# Toward alternative metrics of journal impact: A comparison of download and citation data.

Johan Bollen

*Department of Computer Science, Old Dominion University, 4700 Elkhorn Ave., Norfolk VA 23529*

Herbert Van de Sompel

*Research Library, Los Alamos National Laboratory, Los Alamos, NM, 87554*

Joan A. Smith

*Department of Computer Science, Old Dominion University, 4700 Elkhorn Ave., Norfolk VA 23529*

Rick Luce

*Research Library, Los Alamos National Laboratory, Los Alamos, NM, 87554*

**Abstract**

We generated networks of journal relationships from citation and download data, and determined journal impact rankings from these networks using a set of social network centrality metrics. The resulting journal impact rankings were compared to the ISI IF. Results indicate that, although social network metrics and ISI IF rankings deviate moderately for citation-based journal networks, they differ considerably for journal networks derived from download data. We believe the results represent a unique aspect of general journal impact that is not captured by the ISI IF. These results furthermore raise questions regarding the validity of the ISI IF as the sole assessment of journal impact, and suggest the possibility of devising impact metrics based on usage information in general.

# 1 Introduction

The Institute of Scientific Information's Impact Factor (ISI IF) has served as a de facto definition of the concept of journal impact for the past 40 years. Even today, most research regarding the impact of scholarly publications is focused on the use of citation frequencies, an approach typified by the ISI IF (Garfield, 1979) .

ISI publishes, on a yearly basis, the JCR database (Komatsu, 1996) containing the Impact Factors (IFs) for a core set of about 6000 journals. The ISI IFs are widely regarded as the standard by which to judge the impact of a given journal, the quality of publications for individual authors (Rey-Rocha, Martin-Sempre, & Garzon, 2002), the quality of research for research departments (Jacso, 2000; Kaltenborn & Kuhn, 2003), and the scientific output of entire countries (Kaltenborn & Kuhn, 2003; Bordons, Fernandez, & Gomez, 2002).

However, the general concept of journal impact is a multi-faceted, highly general notion (Rousseau, 2002) which can be defined, operationalized and measured in a number of different ways, and with varying degrees of validity and reliability. We refer to this more general concept of journal impact as a quantifiable entity labeled $I_g$ of which the ISI IF is one representation. Indeed, numerous alternatives to the ISI IF have been proposed on the basis that it does not validly represent journal impact or $I_g$ (Egghe, 1988; Harter & Nisonger, 1997; Nederhof, Luwel, & Moed, 2001; Lewison, 2002).

In this article we propose and examine a set of alternative metrics of journal impact inspired by social network metrics of status. We apply these metrics to journal networks which have been derived from the Journal Citation Records and from sequential journal download patterns registered in the log files of a large Digital Library (DL). The resulting impact rankings can be compared to the ISI IF and reveal different aspects of journal impact both among the global, ISI-defined community of authors and the local community of DL users.

# 2 Background

The concept of journal impact, which we have labeled $I_g$, can be measured and represented in a number of different ways. In the following sections, we examine existing and proposed metrics of journal impact and provide a taxonomy which reveals the lack of journal impact metrics based on usage data, i.e. downloads, and the structural features thereof.

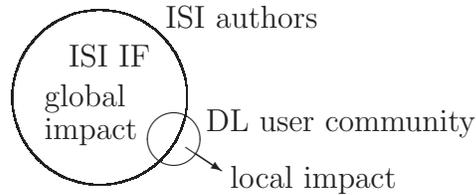

Fig. 1. Local reader and ISI author communities may not overlap entirely.

*2.1 The ISI IF: a frequentist, author generated impact metric*

The ISI IF represents journal impact as the ratio between the number of citations to articles published in a journal over a 2 year period, divided by the total number of citeable articles published in that same period. It expresses the impact or quality of a journal in terms of the degree to which its articles are cited in the literature.

Regardless of the set of assumptions about what motivates authors to cite (MacRoberts & MacRoberts, 1989) that underly the ISI IF, and the issues that arise when it is being applied in a range of domains (Moed & Leeuwen, 1995; Opthof, 1997; Reedijk, 1998), the ISI IF as an operationalization of $I_g$ can be characterized by three main features:

(1) It is based on a frequentist metric: journal impact is largely determined by counting the number of citations to a journal. Each citation is counted as a vote of confidence for the particular journal, and a citation count amounts essentially to a poll of experts (authors) on the impact of a journal.
(2) It is based on a selection made by the ISI of all published journals.
(3) It is determined from citation frequencies as they occur for a global, non-specific community of authors.

First, by its focus on citation frequencies the ISI IF focuses on a highly particular aspect of $I_g$, thereby ignoring more contextual indications of journal impact. For example, do journals which receive citations mostly from high impact journals also have high impact in spite of a relatively low absolute citation count? Does a journal that contains a high number of out-going citations function as a "hub" in the citation graph and thereby have higher impact than the number of its in-coming citations alone would indicate? Does a journal whose articles critically connect different scientific domains have high impact? These examples pertain to structural features of impact which a frequentist metric of $I_g$, such as the ISI IF, does not express.

Second, the ISI IF is calculated on the basis of citation frequencies which have been registered for an ISI-defined set of selected scholarly journals. This core set of journals does not include a majority of the growing body of web-based publications (Groote & Dorsch, 2001; Harnad, Carr, Brody, & Oppenheim, 2003), gray literature (Cesare, 1994), and multimedia collections.

Third, the ISI IF is based on the journal citation patterns of a global community of authors. It thus represents a global, consensus view of journal impact. Local author and reader communities can, however, have strongly diverging views. Therefore the ISI IF, as a "global" metric of impact, cannot provide an accurate assessment of the degree to which a particular DL's collection fits the needs of its local community as shown in among others by (Line, 1977; Bollen, Luce, Vemulapalli, & Xu, 2003). This situation is graphically represented in Fig. 1.

## 2.2 Usage ranking of journals

Since ISI IF journal rankings are not adaptable to specific circumstances, researchers have introduced the notion of applying usage data, such as web hits or downloads, to determine journal rankings.

Kaplan and Nelson (2000) evaluate the impact of a DL by examining journal usage (determined from article downloads) and comparing the resulting rankings to the ISI IF. They conclude that both measures need to be combined to more completely assess the impact or $I_g$ of the set of journals they have included in their analysis.

Similarly, Darmoni, Roussel, Benichou, Thirion, and Pinhas (2002) compare journal usage frequency to the ISI IF for a medical DL collection. They define a "Reading Factor" (RF) which consists of the ratio of a particular journal's download frequency to the total downloads of all journals as recorded in the DL's logs. The authors report a low and statistically insignificant correlation between the observed RF and the ISI IF for the same set of journals. These results show that journal download frequency within a local DL community does not correspond to the ISI IF, which raises questions regarding the ISI IF's validity as the sole indicator of $I_g$ among a specific community of readers.

However, Kurtz et al. (2000, 2005a, 2005b) reports on a comparison of readership and the ISI IF for a set of four high-impact Astrophysics journals. Contrary to (Darmoni et al., 2002), the normative assumption, namely that citation corresponds to readership, is confirmed in these results. In addition, Kurtz et al. (2005a) shows how the obsolescence function (Egghe & Rousseau, 2000) of citations and readership follow similar trajectories across time, and how readership and citation rates can be combined to assess the research productivity of individuals.

Since adequate large-scale statistics on citation frequencies are relatively difficult to freely obtain, (Hitchcock et al., 2002; Harnad et al., 2003) propose to determine document impact from "hit" frequencies in a system of open pre-print archives. In such a system, citation frequencies derived from Open Access publications and "web hit" frequencies derived from DL logs can determine impact and may provide an alternative to the ISI IF.

*2.3 Social network metrics of power and status*

The ISI IF is derived from document citation data which defines a network of journal relationships. Such relationships are determined by counting the frequency by which articles in journal $v_j$ have cited the articles published in journal $v_i$, so that we have a citation frequency count associated with each pair of journals. We can then aggregate these numbers into a network of journal relationships. In such a network the nodes represent individual journals and the edges are assigned weight values according to how frequently articles published in one journal cite articles published in another, as shown in Fig. 2. We can represent such a journal network over $n$ journals by the $n \times n$ journal adjacency matrix $W$ whose entries $w_{ij} \in \mathbb{N}^+$ represent the weight of the relationship between any pair of journals $v_i$ and $v_j$ as they were recorded in the two years preceding year $t$.

The ISI IF can then be redefined as a network metric of node status on the basis of a journal's backlinks in the journal adjacency matrix:

$$IF(v_j, t) = \frac{\sum_{i=1}^{n} w_{ij}}{N} \qquad (1)$$

where $N$ denotes the number of articles published in journal $v_j$.

In other words, the ISI IF represents the normalized graph-theoretical back-link frequency of a journal in a graph of journal citation relationships as shown in Fig. 3. However, for each journal it ignores the large majority of its network relationships, and requires only a journal's backlink frequency. As such it disregards the nature of the backlinking journals and the overall structure of link patterns; it remains a frequentist metric.

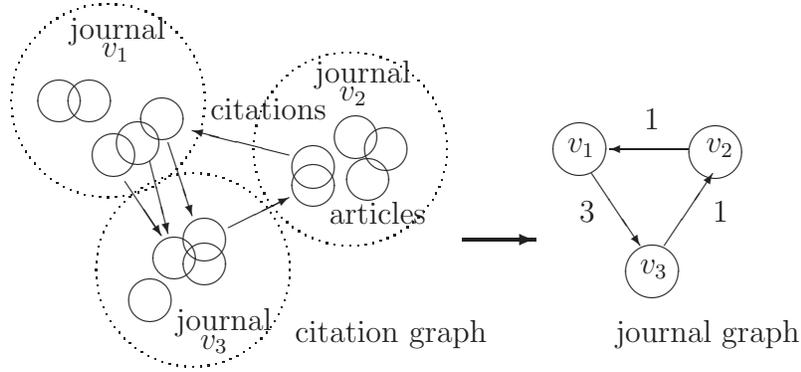

Fig. 2. The article citation graph induces a journal citation graph

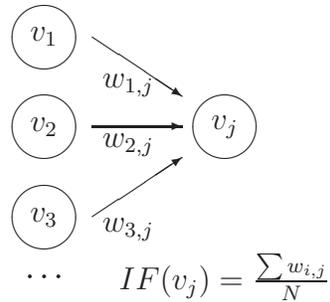

Fig. 3. The ISI IF defined on the basis of a network of journal relationships whose edges represent citation frequencies

The problem of ranking a set of entities according to their status, or impact [1], in a network of relationship data has a long-standing history in sociology, specifically in the domain of social network analysis. These approaches to status ranking go beyond the frequentist approach of the ISI IF by examining the structure of relationships among actors across multiple dimensions (Bonacich, 1987; Wasserman & Faust, 1994).

The three most often applied social network status metrics can be described as follows:

---

[1] We assume status and impact can be used interchangeably with status refering to the more general case of ranking any set of entities and impact specifically refering to information sources such as web pages, articles, journals, etc.

(1) Degree centrality: the sum of the number of relationships pointing to and from an actor, i.e., their in- and out-degree, normalized by the total number of relationships in the social network
(2) Closeness centrality: the average shortest path distance of an actor to all other actors in the network.
(3) Betweenness centrality: the frequency by which an actor is part of the shortest path between any pair of agents in the network.

Fig. 4 shows an example of a directed weighted graph, which may represent a journal citation network, in which the degree, closeness and betweenness centrality of node $v_i$ can be determined. The mentioned centrality metrics are defined such that they accommodate for the presence of weighted, directional citation links. In this case degree centrality is defined as the sum of weights of links pointing to and from any node $v_i$ (Newman, 2004), or:

$$c_d(v_i) = \frac{\sum_j w_{i,j} + \sum_j w_{j,i}}{\sum_{i,j} w_{j,i}}$$

The degree centrality of node $v_i$ in the journal network in Fig. 4 is $c_d(v_i) = w_{i,1} + w_{i,4}$.

Closeness centrality is generally defined as the average weight of the shortest path distance between node $v_i$ and any other node in the network (Wasserman & Faust, 1994). Since this case concerns weighted journal relationships and the weights represent how an author or reader may link one journal to another (flow), we define the shortest path weight between nodes $v_i$ and $v_j$, labeled $w_s(v_i, v_j)$ as the product of its constituent link weights, and the distance between any two nodes consequently as the inverse of their shortest path weight. In Fig. 4 for example, the distance between node $v_i$ and $v_2$, $d(v_i, v_2) = \frac{1}{w_{i,1} \times w_{1,2}}$. We then define:

$$c_c(v_i) = \frac{\sum_j d(v_i, v_j)}{N}$$

where $N$ represents the number of shortest-path connections that exist between any pair of nodes $v_i$ and $v_j$.

As such, the closeness centrality of node $v_i$ in the journal network in Fig. 4 is expressed as:

$$c_c(v_i) = \frac{w_{i,1} + d_{i,2} + d_{i,3} + w_{i,4}}{N}$$

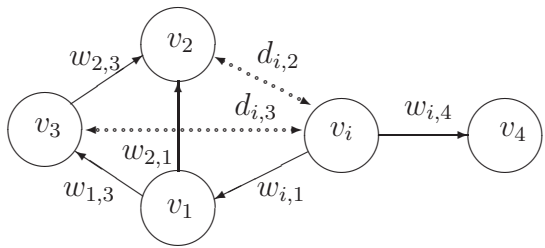

Fig. 4. Example social network.

Betweenness centrality is generally defined as the number of shortest paths that pass through a node $v_i$ (Wasserman & Faust, 1994). Using the above defined shortest path weight, however, we define $v_i$'s betweenness centrality $c_b(v_i)$ as the sum of the weights of the shortest paths that pass through $v_i$:

$$c_b(v_i) = \sum_j w_s(v_i, v_j)$$

In Fig. 4 the betweenness centrality of node $v_i$ would be zero, since it is not on the shortest path connecting any other pair of nodes.

All three centrality metrics represent different aspects of status. Degree centrality, which focuses on the total number of relationships to other nodes can be viewed as an expanded version of the ISI IF, i.e. in addition to a node's (in this case, a journal's) in-degree (back-link frequency), it also takes into account its out-degree. Closeness centrality expands this concept further by not only taking into account the number of immediate neighbors of a node (in- and out-degree), but its network proximity to all other nodes. Betweenness centrality examines how well a node connects pairs of other nodes. A node may have few neighbors and therefore a low degree centrality. However, it may function as a vital bridge through which the paths connecting large groups of nodes pass. In this latter case, its betweenness centrality will reflect this fact.

In the degree, closeness and betweenness centrality metrics we see a gradual expansion of the notion underlying the ISI IF, namely that the status, or rather impact, of a journal can be determined from the number and patterns of its relationships to other journals. Starting from the ISI IF and ending with the betweenness centrality we see a gradually increasing focus on network context and structure, away from frequentist metrics.

Social network and citation analysis have, in the past decade, successfully converged on WWW search engines. Rather than rank web pages according to how strongly their text content matches a particular user query, search engines can identify high-impact pages by examining the context of their hyperlink relationships to other pages. (Kleinberg, 1998, 1999; Brin & Page, 1998). Our efforts to devise an alternative set of journal impact metrics are an attempt to bring the benefits of this approach to the domain of journal impact ranking.

*2.4 A taxonomy of impact measures*

When we examine existing systems for impact ranking we observe that they generally differ in terms of two dimensions: the impact metrics employed, and the data sets they have been applied to. The first dimension corresponds to whether an impact measure is based on frequency-based metrics as opposed to structural metrics. The second dimension concerns whether a measure is based on author- or reader-defined data sets. Most of the discussed impact measures vary on both dimensions and can be classified accordingly, although some measures combine features of both dimensions. Fig. 5 represents an overview of this classification.

Because the ISI IF combines a frequency-based impact metric with an author-defined data set (citations), we place it in the upper-left quadrant. The Reading Factor (RF) as proposed by Darmoni et al. (2000) can be situated in the upper-right quadrant of Fig. 5. Its data set is based on the preferences of a specific community of readers, i.e. article downloads, but does not take into account any of the structural features associated with such preferences. The RF impact metric, based solely on frequency counts, essentially amounts to a reader-defined version of the ISI IF.

The bottom right and left quadrants are associated with impact metrics that rely on the structural features of a set of relationships among documents, journals, web pages, etc. These metrics can be applied to reader-generated data vs. author-generated data, and are, as such, respectively labeled Structural-Author (SA) and Structural-Reader (SR).

The SA metrics determine the impact of documents, web pages, journals, etc. on the basis of the structure of a network of authored relationships. A majority of this work has focused on the impact ranking of web pages, but is equally applicable to citation graphs. A prominent example of this approach is the Google search engine (Brin & Page, 1998) which determines a web page's impact on the basis of its hyperlink relationships to other web pages by means of the PageRank metric. The PageRank rankings of Google have also been used to define a Web Impact Factor by Thelwall (2001). Other approaches are based on the same principle but decompose the concept of impact into a level of "hubness" and "authorativeness", e.g. the HITS system (Chakrabarti et al., 1998; Kleinberg, 1999).

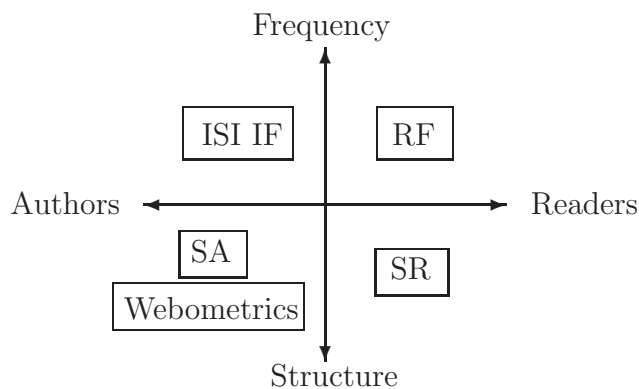

Fig. 5. Frequency vs. Structure and Author vs. Reader impact metrics

Given the limitations of static, author-designed networks, some systems attempt to merge features of the SA and SR metrics by introducing usage data in the ranking of web pages. Zhu, Hong, and Hughes (2001) discuss combining Google's PageRank with user ratings and hit rates. Nevertheless, little research has been conducted in the area of the fourth quadrant, i.e., a combination of structural impact metrics and reader-defined data sets (marked "SR" in Fig. 5). We speculate this can be attributed to the lack of methodologies to efficiently extract reader generated networks. However, the SR quadrant represents a promising combination of network-based metrics of impact and data sets based on the preferences and views of a community of readers.

2.5 *The SR Quadrant: reader-based structural metrics of journal impact*

Our research represents a synthesis of structural and reader-based approaches (the SR quadrant) to determine an impact ranking of journals. Rather than determining a journal's impact from the frequency with which its articles are being cited, we take into account its topological context within a network of journal relationships. These networks are determined not by author citations but by the full-text article downloads of a specific community of readers within a DL.

We developed a methodology to construct networks of journal relationships using reader download patterns, and applied a set of social network metrics to them to determine alternative journal impact rankings. These impact measures would be situated in the lower-right quadrant of the taxonomy outlined

| Community | | Data | | Journal Relationships | | Impact Metric |
|---|---|---|---|---|---|---|
| Authors | → | Citation data | → | AGN | → | Author Journal Impact |
| Readers | → | Download data | → | RGN | → | Reader Journal Impact |

Table 1

Reader generated impact metrics are generated via a process analogous to traditional citation analysis.

in Fig. 5: a data set based on usage, approximated by downloads, and a metric based on structural network features.

A schematic overview of this approach is visualized in Table. 1. We start from two communities which we represent separately, although in reality they overlap to some degree, namely readers and authors.

As authors cite articles, they thereby collectively define an aggregate set of article citation data. This citation data set can be used to generate sets of journal relationships such as the ISI JCR database, i.e. how often do articles in journal A cite articles published in journal B. From such citation data, one can derive journal relationship networks which we will label Author Generated Networks (AGN). Author-determined journal impact metrics such as the ISI IF are then calculated on the basis of such data.

Analogous to this process we propose to derive reader-determined journal metrics from the journal relationships expressed in reader download patterns in a DL. We start with a community of readers who download full-text articles in a DL, analogous to a community of authors who cite articles from their publications. In this case, usage, as indicated by download data, replaces citation as a definition of journal relationships. More precisely, rather than determine how often authors have cited articles in journal B from articles published in journal A, we examine how often articles in journal B have been downloaded within the same session as articles in journal A. Subsequently, networks of journal relationships can be derived from such download data. The resulting Reader Generated Networks (RGN) constitute an alternative representation of journal relationships and can be juxtaposed with the AGN since they represent journal relationships defined by the actions of local readers, rather than by the citations of authors as registered by the ISI. Structural metrics of journal impact can be defined on the basis of these reader-determined networks of journal relationships.

## 2.6 Research Questions

This article concerns three particular research questions:

**Question 1** Can valid networks of journal relationships be derived from reader article download patterns registered in a DL's server logs?

**Question 2** Can social network metrics of journal impact validly be calculated from the structure of such networks?

**Question 3** If so, how do the resulting journal impact rankings relate to the ISI IF?

We address these issues by following the specific steps outlined below.

First, regarding **question 1**, we have generated an RGN from reader article download data. We provide support for its validity by first examining its small-world characteristics such as the average node distance compared to those given random graph models. We proceed with an expert evaluation of journal relationships: a team of 22 LANL scientists rate individual journal-to-journal relationships as given by the RGN. We finally correlate the RGN and AGN journal relationship weights to determine whether the RGN matches the citation-based AGN.

Second, regarding **question 2**, we validate the social network metrics by calculating them over the structure of an AGN derived from the ISI JCR data set, and comparing the results to the ISI IF rankings. Since both are calculated from citation data, the resulting journal rankings should be comparable. If the social network metrics generate a journal impact ranking comparable to the ISI IF, this validates their ability to express $I_g$ on the condition one supports the assumption that the ISI IF is a valid expression of $I_g$.

Third, regarding **quesstion 3**, we examine social network metrics calculated on the basis of both the AGN and RGN. By comparing the resulting journal impact values, we determine whether these differ strongly when generated for a community of authors vs. readers. In addition, we compare AGN and RGN structural impact rankings to the ISI IF to determine how the ISI IF as an impact metric compares to the applied structural metrics.

# 3 Constructing Reader Generated Networks

## 3.1 Retrieval Coherence Assumption

The main principle underlying the generation of the RGN is the Retrieval Coherence Assumption (RCA), namely the notion that when a DL user downloads a set of documents he or she is often driven by a specific information need. From the RCA it follows that when we observe a reader sequentially downloading a set of articles, we can infer a certain probability that the downloaded articles, and thus the journals in which they appeared, are related. Their degree of relatedness can be determined on the basis of two factors. First, the closer the documents are located within a sequence of reader downloads, the more related they are expected to be (Jones, Cunningham, & McNab, 1998; Pirolli & Pitkow, 1999; Chi, Pirolli, & Pitkow, 2000). The RCA thus applies most reliably to the shortest retrieval sequences, i.e. pairs of documents downloaded one after the other. Second, the more frequently a particular pair of documents are downloaded by a group of readers, the greater the degree to which we can assume these documents to be related.

For example, when readers frequently download article A shortly after downloading article B, this may indicate A and B are related by a common user information need. Consequently, the journal in which A was published may be related to the journal in which B was published. This download sequence of two documents thus reveals implicitly whether or not the two journals in which A and B were published are related in the reader's mind. Given that we have a DL log which records a large set of document downloads, we can reconstruct reader download sequences and use these to determine journal relationships. Such an approach is strongly related to item-based collaborative filtering techniques (Sarwar, Karypis, Konstan, & Reidl, 2001; Huang, Chen, & Zeng, 2004), market basket analysis (Brin, Motwani, & Silverstein, 1997), and clickstream data mining (Yan, Jacobsen, Garcia-Molina, & Dayal, 1996; Mathe & Chen, 1996; Xiao & Dunham, 2001) which analyse user downloads and hyperlink traversals to generate a set of document relationships.

We have applied the RCA to the construction of the RGN. An algorithm updates journal relationship weights according to the frequency with which articles within these journals have been downloaded in temporal proximity. The download patterns of individual readers each contribute small amounts to journal relationship weights, and induce only small changes in the RGN. However, consistent download patterns over a group of readers will gradually establish a set of significant journal relationship weights which reflect the degree to which articles within the two journals have been downloaded by readers within the same download sequence.

3.2 *Reconstruction of downloads sequences from DL Logs*

The Los Alamos National Laboratory (LANL) Research Library (RL) was used as a test-bed for the generation of the RGN. A large portion of the RL collection is available in digital, full-text format. Readers can download most journal articles to their desktops from the RL's web site. RL logs were extracted for the months June to November of the year 2001. These logs registered 40,847 full-text article downloads for 1,858 unique users. Although the LANL RL offers access to at least 20,000 journals the downloaded articles only spanned a range of 1,892 journals.

The RL log files contain an IP address, date and time of the download, and a document identification for each downloaded article. The IP addresses will be used to identify individual users. Although an IP address is not always a unique user identifier due to proxies, most LANL RL users have their own machines which each have unique IP addresses. Each article was identified by a SICI identification number which contained the ISSN (International Standard Serial Number: a universal journal identification number) of the journal in which the article appeared.

The LANL RL logs were processed as follows. First, to protect the privacy of individual users, yet retain information on where a specific request originated from, we replaced all IP addresses with an anonymized, unique user identifier (ID). Second, we extracted the ISSN number from each document SICI and added it to the DL log so that an ordered sequence of journal ISSNs could be reconstructed for each user ID. Third, all registered retrieval events in the DL log were ordered according to user ID and retrieval time.

We then examined every two subsequent lines in the resulting file according to the following criteria:

(1) Were the article download requests issued by the same user, i.e. same user IDs?
(2) Did the article download time-stamps differ by less than one hour? ($\Delta_t$ = 3600 seconds)

If any pair of download requests were found to conform to these criteria, i.e. they were issued by the same user within the same session, they were assumed to be related according to the RCA. According to the RCA, we can define such a pair of downloads as a co-retrieval event.

| user ID | date/time | Document ID | ISSN | latency (s) |
|---|---|---|---|---|
| 100 | 2001-08-24T17:12:52-05:00 | 02721716;14;4;69_ddacfrtip | 0272-1716 | - |
| 100 | 2001-08-24T17:14:41-05:00 | 01689274;25;4;499_prtuauma | 0168-9274 | 109 |
| 100 | 2001-08-24T17:15:43-05:00 | 00978493;19;2;281_apiaaaortars | 0097-8493 | 62 |
| 101 | 2001-06-18T12:03:04-05:00 | 00207225;38;3;347_otfrim | 0020-7225 | - |
| 101 | 2001-06-18T12:04:40-05:00 | 02780062;19;3;211_aotdfmfct | 0278-0062 | 96 |
| 101 | 2001-06-18T13:13:40-05:00 | 08956111;25;2;113_asrgt3tr | 0895-6111 | 3140 |
| ↓ | coretrievals ↓ | | ↓ | $< \Delta_t$ |
| 100: | 02721716;14;4;69_ddacfrtip→01689274;25;4;499_prtuauma | | 0272-1716→ 0168-9274 | |
| 100: | 01689274;25;4;499_prtuauma→00978493;19;2;281_apiaaaortars | | 0168-9274→ 0097-8493 | |
| 101: | 00207225;38;3;347_otfrim→02780062;19;3;211_aotdfmfct | | 0020-7225 → 0278-0062 | |
| 101: | 02780062;19;3;211_aotdfmfct→08956111;25;2;113_asrgt3tr | | 0278-0062 → 0895-6111 | |

Table 2
User downloads in a DL log indicate reconstruct document and journal co-retrieval events.

An example of this procedure is shown in Table 2 which lists a sequence of document downloads as they were registered for the Los Alamos National Laboratory RL services. For example, user 100 downloads two consecutive documents, e.g. "02721716;14;4;69_ddacfrtip" and "01689274;25;4;499_prtuauma". Since these documents have been downloaded on the same date and within only 109s, i.e. the download latency is less than the given $\Delta_t$ threshold of 3600s, the two downloads constitute a co-retrieval event. The relationship between the two journals involved in the document co-retrieval event can thus be updated.

Once a set of co-retrieval events has been reconstructed from a DL log they can be used to generate a journal network (RGN). We define an algorithm which increases the relationship weight between any two journals involved in a co-retrieval event as follows:

Given that the DL log from which we have distilled a set of co-retrieval events concerns a set of $n$ journals, we can represent the RGN by the $n \times n$ matrix $\mathbb{R}$ whose entries $r_{ij} \in \mathbb{N}^+$ represent the strength of journal relationships between any pair of journals $v_i$ and $v_j$.

We denote the set of download sequences derived from a given DL log as $E = \{e_1, e_2, \cdots, e_k\}$. Each co-retrieval event $e_i$ is represented by the triplet $e_i = (v_i, v_j, t(v_i, v_j))$ where $t(v_i, v_j)$ represents the time in seconds elapsed between the downloads of document $v_i$ and $v_j$, and $t(v_i, v_j) < \Delta_t$. Each co-retrieval for the documents $v_i$ and $v_j$ corresponds to a small increase of journal relationship weight, $\rho$, which is added to $r_{ij}$ and defined as $\rho = f(e_i)$. The reinforcement function $f(e_i)$ can be varied according to the nature of the data set on which the algorithm is operating.

For all applications discussed in this article, we will define $f(e_i) = 1$ which means $r_{ij}$ will correspond exactly to the frequency with which a given co-retrieval will occur over the downloads registered in a DL log.

We can then formalize the algorithm to produce the weight values for matrix $R$ as follows:

$\forall_{ij} a_{ij} = 0$
**for** (i=1; i<n+1; i++){
  $e_i = (v_i, v_j, t(v_i, v_j))$: $r_{ij} + = f(e_i)$
}

Every co-retrieval $c_i = (v_i, v_j, t(v_i, v_j))$ corresponds to a small reinforcement value $\rho$ added to the matrix entry $r_{ij}$ which represents the strength of the relationship between the journals $v_i$ and $v_j$. In this sense, the set of all overlapping "trails" of co-retrieval events gradually generates a journal network. This network can be held to represent the preferences of the user community for which the set of co-retrievals has been generated. Previous experiments demonstrate how such networks represent the collective views and preferences of a specific community of readers (Bollen, 2001).

The proposed methodology can be applied similarly to the construction of article networks. In fact, DL logs naturally store download data on the article level which we translated to the journal level. A DL log analysis could thus yield both journal and article RGNs which could be analysed separately to determine article as well as journal relationships, and associated impact rankings.

### 3.3 General RGN features

An RGN was generated for 1892 journals, represented by a 1892 × 1892 matrix of journal relationship weights. Table 3 shows an extract of matrix $R$ for a subset of 10 journals. Journal ISSN numbers were converted to the abbreviated journal titles for readability. The entries of this matrix correspond to the weight values of journal relationships in the RGN. As shown, the journals "Physica B" and "Physica C" are connected by a journal relationship whose weight is 18, which indicates a relatively powerful relationship compared to other journal relationships such as those between the journals "Journal of Magnetism and Magnetic Matter" and "Physica B" (weight 11).

| Journal Title | Index | 01 | 02 | 03 | 04 | 05 | 06 | 07 | 08 | 09 | 10 |
|---|---|---|---|---|---|---|---|---|---|---|---|
| J COMPUT PHYS | 01 | 0.0 | 0.0 | 0.0 | 0.0 | 0.0 | 0.0 | 0.0 | 0.0 | 0.0 | 0.0 |
| PHYSICA B | 02 | 1.0 | 0.0 | 0.0 | 1.0 | 18.0 | 1.0 | 0.0 | 6.0 | 0.0 | 12.0 |
| INT J COMPUT VISION | 03 | 0.0 | 0.0 | 0.0 | 0.0 | 0.0 | 0.0 | 0.0 | 0.0 | 0.0 | 0.0 |
| NUCL INSTRUM METH A | 04 | 0.0 | 1.0 | 0.0 | 0.0 | 0.0 | 0.0 | 0.0 | 0.0 | 0.0 | 0.0 |
| PHYSICA C | 05 | 0.0 | 22.0 | 0.0 | 0.0 | 0.0 | 0.0 | 0.0 | 0.0 | 0.0 | 1.0 |
| MAT SCI ENG A-STRUCT | 06 | 0.0 | 0.0 | 0.0 | 0.0 | 0.0 | 0.0 | 6.0 | 2.0 | 0.0 | 0.0 |
| ACTA MATER | 07 | 0.0 | 0.0 | 0.0 | 0.0 | 0.0 | 10.0 | 0.0 | 0.0 | 0.0 | 0.0 |
| J ALLOY COMPD | 08 | 0.0 | 7.0 | 0.0 | 0.0 | 1.0 | 3.0 | 0.0 | 0.0 | 0.0 | 3.0 |
| J POWER SOURCES | 09 | 0.0 | 0.0 | 0.0 | 0.0 | 0.0 | 0.0 | 0.0 | 0.0 | 0.0 | 0.0 |
| J MAGN MAGN MATER | 10 | 0.0 | 11.0 | 0.0 | 0.0 | 2.0 | 2.0 | 0.0 | 4.0 | 0.0 | 0.0 |

Table 3
Sample of generated RGN matrix for 15 journals from which articles were most frequently downloaded.

As expected, matrix $R$ was sparse: the ratio of non-zero entries to the total number of matrix entries ($1892^2 - 1892$) was 0.176%. Indeed, only a small fraction of all possible, directed journal relationships can be meaningful and therefore matrix densities will be low. The distribution of values in matrix $R$ indicated a wide range of possible journal relationship weights: the mean of link weights was found to be 1.195 with a standard deviation of 0.821 for all $r_{ij} : r_{ij} > 0$. The minimum and maximum recorded values over all non-zero entries were found to be 1 and 22 respectively, indicating user co-retrieval events strongly focused on specific pairs of journals.

## 4   Author Generated Networks

After generating an RGN on the basis of DL download data, we generate an AGN on the basis of citation data derived from the JCR database. The JCR database contains the ISI IF in conjunction with raw journal citation counts, i.e. the number of citations that occur from articles published in one journal to those published in another. These citation counts are split according to publication year so that citation counts among any pair of journals can be retrieved for specific years.

To allow comparisons between the generated AGN network and the previously generated RGN network, we extracted citation counts only for those journals which occurred in the log data used to generate the RGN. Both networks would as such pertain to the same set of 1892 journals. Citation counts for this set of 1892 journals were extracted from the 2001 JCR database which was recorded for the 2 years preceding 2001, namely 2000 and 1999.

We represent the AGN by the $n \times n$ matrix $A$ whose entries $a_{ij} \in \mathbb{N}^+$ represent the sum of the number of citations between two journals, i.e. $v_i, v_j \in V = \{v_0, v_1, \ldots v_n\}$ ($V$ represents the set of RGN and AGN journals). Such citation counts are extracted from the JCR database for the 2 years preceding 2001. For example, if $a_{ij} = 4$, this indicates the articles published in journal $v_i$ cited articles published in journal $v_j$ 4 times in the period 1999 and 2000.

Matrix $A$ was more dense than matrix $R$. 36,617 non-zero entries were found over a total number of $1892^2 - 1829$ possible entries, bringing matrix density to 1.0946%. Network connection weights for all $a_{ij} : a_{ij} > 0$ ranged from 0 to 3735, with a mean of 12.5648 and a standard deviation of 70.0512.

## 5 RGN Network Validation

We have at this point generated an RGN and AGN journal network represented by two matrices, $R$ and $A$. The AGN is based on the 2001 JCR data, but the RGN's validity as a meaningful journal-relationship network needs to be confirmed in order to extract usage-based metrics of journal impact from the RGN. Although no formal proof of RGN validity can be generated, the following three criteria may provide strong support for its validity as a representation of local LANL journal relationships:

(1) Does the general network structure exhibit the features of a small world network?
(2) Does the RGN validly represent the views and preferences of the local LANL community?
(3) What is the degree of similarity between the generated RGN and AGN?

### 5.1 Small world network features

The distribution of co-retrieval frequencies was highly skewed: the most frequent co-retrieval event had a frequency of 22, while 5250 co-retrievals oc-curred only once. Table 4 list the 5 highest valued journal relationships in the RGN.

| Start node | End Node | Frequency |
| :---: | :---: | :---: |
| PHYSICA C | PHYSICA B | 22 |
| PHYSICA B | PHYSICA C | 18 |
| NUCL INSTRUM METH A | IEEE T NUCL SCI | 14 |
| PHYSICA B | J MAGN MAGN MATER | 12 |
| J MAGN MAGN MATER | PHYSICA B | 11 |

Table 4
Five pairs of journals for which highest co-retrieval frequencies have been found.

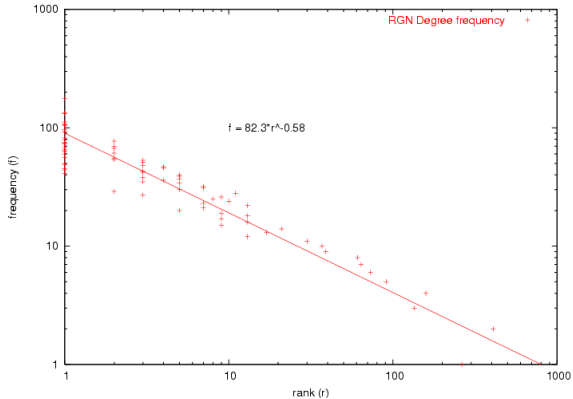

Fig. 6. Journal degree frequency closely follows a powerlaw function in the RGN.

Co-retrieval frequency and rank thus closely fit an inverse power function $f = 5250r^{-3.6}$. As expected, this result indicates co-retrieval event frequencies follow a similar pattern to that found for WWW page retrievals (Breslau, Cao, Fan, Phillips, & Shenker, 1999; Levene, Borges, & Loizou, 2001). In addition we find that the degree frequencies for journals in the RGN closely follow a power law as shown in Fig. 6. The relationship between journal degree frequency ($f$) and rank ($r$) can be fit by the function $f = 82.3r^{-0.58}$, and is a strong indication of this network's scale-free topology (Barabasi & Albert, 1999; Watts, 1999). Given that the mean out-degree of journals in the RGN network is 3.7, we can expect an average node distance of ±5.76. However, the actual mean node distance of the largest component in the RGN network is significantly less, namely ±4.6, adding further support to the assumption that the RGN is a small-world graph (Newman, 2000).

| rating | frequency |
|---|---|
| very strong | 16% |
| strong | 18% |
| moderate | 32% |
| weak | 26% |
| very weak | 8% |
| total: | 100% |

Table 5
Frequency table for LANL scientist evaluations of journal relationships in RGN

*5.2 LANL expert evaluation of RGN*

A sample of the RGN network was validated by a team of 22 LANL scientists. These scientists were selected on the basis of whether they had published in any of the 10 most frequently downloaded journals in the past 3 years. The ten selected journals are listed in Table 3.

Each scientist was invited by email to rate the RGN relationships between the journal he or she published in (the cue journal) and at the most ten other journals (the target journals). All ratings were collected via a web form. This web form listed the cue journal followed by a list of target journals. The strength of relationship between the cue and each of the target journals could be rated on a 5 point scale, namely "very strong", "strong", "moderate", "weak", "very weak" or "no opinion". Each scientist was reminded to not rate the journals themselves but the relationships between the cue and target journals.

204 journal to journal evaluations were obtained, of which 38 indicated "no opinion". The latter were treated as missing data and excluded from subsequent analysis. The mean rating for all journal relationships was 2.9398 , with a median of 3 and a standard deviation of 1.8 corresponding to a rating of "moderate" to "very strong" for 66% of the evaluated journal relationships. Table 5 provides a breakdown of the rating frequencies.

The selected scientists were found to be highly critical in their evaluation of journal relationships. In quite a number of cases the relationship between journals such as "Acta Materialia' and "International Journal of Fracture" would be rated "very weak" even though both correspond to related subjects in material science. The number of "moderate" to "very strong" ratings should therefore be considered very significant.

The LANL scientist ratings reflect a ranking of journal relationships which may or may not correspond to the ranking resulting from the RGN weight values. Clearly, a weakly valued RGN relationship should correspond to a moderate to very weak scientist evaluation and vice versa. Indeed, the Spearman rank correlation between journal relationship weights in the RGN and the scientist evaluations was found to be $\rho = 0.31$ ($p < 0.01$, df=165) indicating that where "moderate" to "very weak" evaluations occurred these corresponded to low-valued RGN rankings and vice versa.

These results indicate that our sample of LANL scientists agree to a large extent with the journal relationships recorded in the RGN, and hence support the validity of the RGN network that was extracted from LANL download data.

### 5.3 AGN and RGN comparison

Journal citations will to some measure correspond to journal download data since the latter is a requirement for the former in the research-publication cycle. A certain degree of similarity between the AGN and RGN journal relationships would thus be expected. To compare journal relationships across the AGN and RGN networks, we calculated the Spearman rank correlation coefficient across the journal relationship weights in both RGN and AGN journal networks. Considering that link weights of zero in the RGN network can indicate either an absence of sufficient log data, or an actual absence of a journal relationship of a significant weight, we calculated the Spearman rank correlation coefficient for all non-zero entries. It was found to be $0.243261$ ($p < 0.01$).

This moderate correlation indicates that relationship weights across RGN and AGN journal networks in part represent a similar structure of journal relationships, but each contain specific and unique information on the relationships among journals uniquely determined by the preferences of LANL readers and ISI authors.

## 6 Journal impact from download and citation data

Journal centrality values were calculated and compared for all journals in the generated RGN and AGN networks, and finally compared to the ISI IF.

| IF | Journal Title |
|---|---|
| 29.219 | Cell |
| 21.568 | Curr. Opin. Cell Bio. |
| 18.866 | Immunity |
| 18.135 | Trends Cell Biol. |
| 16.475 | Trends Neurosci. |
| 14.329 | Trends Biochem. Sci. |
| 14.153 | Neuron |
| 14.091 | Surf. Sci. Rep. |
| 14.000 | Prog. Mater. Sci. |
| 13.724 | Curr. Opin. Immunol. |

Table 6
ISI IF ranking of journals most frequently read in Los Alamos National Laboratory.

The objective of this analysis was to compare three impact rankings of our set of 1892 journals: centrality values calculated for the citation-based AGN journal network, centrality values calculated for the local, LANL reader-based RGN journal network, and finally the ISI IF. We examined the effects of using different centrality metrics for AGN and RGN, namely degree, closeness and betweenness centrality, and interpret the resulting journal impact rankings. An overview of the results is shown in Tables 9, 10, and 11.

### 6.1 ISI IF ranking for LANL journals

We retrieved ISI IF values for the set of 1892 journals used to generate the RGN and AGN, i.e. the set of journals that appeared in the LANL RL download logs. A sample of how the ISI IF ranks this set of journals (10 highest scoring) is shown in Table 6.

We note a preponderance of journals associated with cell biology and immunology, e.g. "Cell", "Curr. Opin. Cell. Bio", and "Immunity". From this list one would conclude that among the journals used at LANL, these would be the highest impact ones and therefore most important to the LANL community. However, although such research topics are addressed at LANL, they hardly represent its strong focus on physics and nuclear science.

|   | AGN Journal Centrality Metrics | | | |
|---|---|---|---|---|
| Rank | Degree (AGN $c_d$) | Closeness (AGN $c_c$) | Betweenness (AGN $c_b$) | ISI IF |
| 1 | Phys. Letter. B | Chem. Phys Lett. | P. IEEE | Cell |
| 2 | Angew. Chem. Int. EditA | J. Colloid. Interf. Sci. | Neural Networks | Curr. Opin. Cell Bio. |
| 3 | Nucl. Phys. B | Angew. Chem. Int. Edit | J. Theor. Biol. | Immunity |
| 4 | Tetrahedon Letter. | J. Chromatogr. | Physica D | Trends Cell Biol. |
| 5 | Lancet | Anal. Chim. Acta. | Chem. Geol. | Trends Neurosci. |
| 6 | Cell | Febs. Lett. | Rep. Prog. Phys. | Trends Biochem. Sci. |
| 7 | J. Mill. Biol. | J. Phys.-Condens. | Curr. Opin. Biotech. | Neuron |
| 8 | Nucl. Inst. Meth. | J. Mol. Biol. | Curr. Opin. Chem. Biol. | Surf. Sci. Rep. |
| 9 | Chem. Phys. Lett. | J. Phys. A-Math. Gen. | Biosens. Bioelectron. | Prog. Mater. Sci. |
| 10 | Febs. Lett. | Anal. Biochem. | Geochim. Cosmochim. Ac. | Curr. Opin. Immunol. |

| Correlation | |
|---|---|
| AGN $c_d$ - ISI IF | 0.408 |
| AGN $c_c$ - ISI IF | 0.147 |
| AGN $c_b$ - ISI IF | 0.131 |

Table 7
AGN rankings for ten highest scoring degree, closeness and betweenness centrality journals compared to ISI IF.

*6.2 Journal impact ranking: AGN centrality values*

First, we verify whether the journal centrality metrics calculated from the AGN structure correlate with the the ISI IF. This comparison serves as a validation of the degree, closeness and betweenness centrality metrics: since the AGN is derived from the same citation counts as the ISI IF, AGN centrality metrics are expected to correlate with the ISI IF and produce similar journal impact rankings. In particular, the AGN degree centrality (AGN $c_d$) is expected to correlate strongly with the ISI IF. It is based on the same frequentist analysis of the citation graph (JCR): its definition entails the summation of the in- and out-degree of journals and thereby partially corresponds to the in-degree data on which the IF is based.

Table 7 shows the impact ranking of journals that results from the degree, closeness and betweenness centralities calculated from the AGN. Only the 10 highest-scoring journals are listed for each. The impact ranking of journals according to the ISI IF is listed in the third, right-most column. The degree centrality ranking for the AGN, even though based on JCR data (like the ISI IF), seems to correspond better to the LANL central mission in nuclear science. We find journals such as "Phys. Letter B', "Nucl. Phys. B" and "Nucl. Inst. Meth. A" ranked among the set of ten highest-valued degree centrality scoring journals. A Spearman rank correlation coefficient indicates AGN degree centrality and the ISI IF overlap to a significant degree, namely 0.408

($p < 0.01$).

The closeness and betweenness centrality values show a striking deviation from the ISI IF rankings. The closeness centrality ranking seems to favor chemistry and molecular biology. The betweenness centrality seems to particularly focus on complexity, biotechnology and biology. Betweenness centrality indeed ranks journals according to how well they bridge the connections between clusters of other journals. Therefore the resulting betweenness rankings may indicate the "connective" role that biology and biotechnology plays in the community of ISI authors. Since the AGN was generated on the basis of ISI citation data, it does not necessarily reflect the views of the local LANL community, nor does the betweenness centrality ranking. The correlations between the closeness and betweenness centrality values and the ISI IF are low but significant, respectively 0.147 ($p < 0.01$), and 0.131 ($p < 0.01$).

*6.3 Journal impact ranking: RGN centrality values*

The above mentioned correlations between the AGN centrality metrics and the ISI IF confirm the ability of the applied centrality metrics to serve as alternative indications of journal impact. We proceed to investigate the journal impact rankings that result from applying the mentioned centrality metrics to the RGN, i.e. the local, LANL-specific network of journal relationships.

Table 8 shows the impact rankings resulting from the degree, closeness and betweenness centrality calculated on the basis of the RGN. For all centrality metrics, we find a journal ranking that deviates strongly from the ISI IF, but reflects features of the local LANL community. In particular, the degree centrality ranking has a stronger focus on nuclear science and material science than both the ISI IF and the AGN degree centrality ranking, as evidenced by the presence of "Physica B" (condensed matter), "IEEE Nucl. Sci", "J. Nucl. Mater.", and "Mat. Sci. Eng A". The correlation of the RGN degree centrality with the ISI IF was non-significant at 0.075, indicating the two rankings are not related.

The closeness centrality rankings focus more strongly on nuclear science and theoretical physics indicating the central position these subjects hold for the local LANL community. The betweenness centrality, on the other hand, favors microbiology and analytical chemistry. This may indicate these journals serve to bridge interests of LANL RL readers, pointing at possible future research venues that were emerging when these logs were recorded and are related to RGN rankings for ten highest scoring degree, closeness and betweenness centrality journals compared to ISI IF.

| | RGN Journal Centrality Metrics | | | |
|---|---|---|---|---|
| Rank | Degree (RGN $c_d$) | Closeness (RGN $c_c$) | Betweenness (RGN $c_b$) | ISI IF |
| 1 | Physica B | IEEE T. Nucl. Sci. | J. Chromatogr. A | Cell |
| 2 | Nucl. Instrum. Meth. A. | Physica B | FEMS Microbiol Lett. | Curr. Opin. Cell Bio. |
| 3 | Mat. Sci. Eng. A-Struct | J. Radioanal Nucl. Ch. | J. Phys. D Appl. Phys. | Immunity |
| 4 | J. Alloy Compd. | J. Comput. Phys. | Talanta | Trends Cell Biol. |
| 5 | IEEE T. Nucl. Sci. | Phys. Lett. A | J. Pharmaceut. Biomed. | Trends Neurosci. |
| 6 | Catal Today | Appl. Surf. Sci. | Photosynth. Res. | Trends Biochem. Sci. |
| 7 | J. Catal. | J. Nucl. Mater. | Electron Lett. | Neuron |
| 8 | Surf. Sci. | Surf. Sci. | Meas. Sci. Technol. | Surf. Sci. Rep. |
| 9 | J. Nucl. Mater. | Nucl. Instrum. Meth. A | Chem. Phys. Lett. | Prog. Mater. Sci. |
| 10 | Appl. Surf. Sci . | Adv. Space Res. | Forensic Sci. Int. | Curr. Opin. Immunol. |

| Correlation | |
|---|---|
| RGN $c_d$ - ISI IF | 0.075 |
| RGN $c_c$ - ISI IF | 0.058 |
| RGN $c_b$ - ISI IF | 0.030 |

Table 8

DEGREE CENTRALITY AND ISI IF

| Rank | AGN Degree Centrality (AGN $c_d$) | RGN Degree Centrality (RGN $c_d$) | ISI IF |
|---|---|---|---|
| 1 | Phys. Letter. B | Physica B | Cell |
| 2 | Angew. Chem. Int. Edit | Nucl. Instrum. Meth. A. | Curr. Opin. Cell Biol. |
| 3 | Nucl. Phys. B | Mat. Sci. Eng. A-Struct | Immunity |
| 4 | Tetrahedon Letter. | J. Alloy Compd. | Trends Cell. Biol. |
| 5 | Lancet | IEEE T. Nucl. Sci. | Trends Neurosci. |
| 6 | Cell | Catal Today | Trends Biochem. Sci. |
| 7 | J. Mill. Biol. | J. Catal. | Neuron |
| 8 | Nucl. Inst. Meth. A | Surf. Sci. | Surf. Sci. Rep. |
| 9 | Chem. Phys. Lett. | J. Nucl. Mater. | Prog. Mater. Sci. |
| 10 | Febs. Lett. | Appl. Surf. Sci . | Curr. Opinion Immunol. |

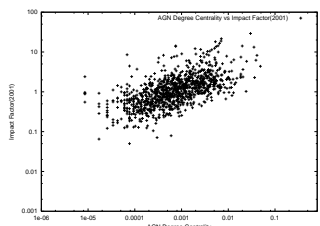 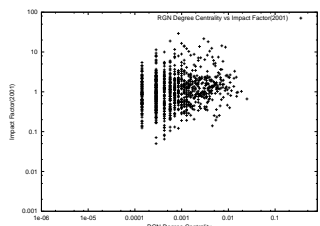 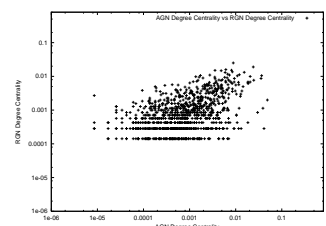

| AGN $c_d$ - ISI IF | RGN $c_d$ - ISI IF | AGN $c_d$ - RGN $c_d$ |
|---|---|---|
| r=0.408 ($p < 0.01$) | r=0.075 | r=0.413 ($p < 0.01$) |

Table 9

Comparison of degree centrality for AGN and RGN, and the ISI IF

| Rank | AGN Closeness Centrality (AGN $c_c$) | RGN Closeness Centrality (RGN $c_c$) | ISI IF |
|---|---|---|---|
| 1 | Chem. Phys Lett. | IEEE T. Nucl. Sci. | Cell |
| 2 | J. Colloid. Interf. Sci. | Physica B | Curr. Opin. Cell Biol. |
| 3 | Angew. Chem. Int. Edit | J. Radioanal Nucl. Ch. | Immunity |
| 4 | J. Chromatogr. A. | J. Comput. Phys. | Trends Cell. Biol. |
| 5 | Anal. Chim. Acta. | Phys. Lett. A | Trends Neurosci. |
| 6 | Febs. Lett. | Appl. Surf. Sci. | Trends Biochem. Sci. |
| 7 | J. Phys.-Condens. Mat. | J. Nucl. Mater. | Neuron |
| 8 | J. Mol. Biol. | Surf. Sci. | Surf. Sci. Rep. |
| 9 | J. Phys. A-Math. Gen. | Nucl. Instrum. Meth. A | Prog. Mater. Sci. |
| 10 | Anal. Biochem. | Adv. Space Res. | Curr. Opinion Immunol. |

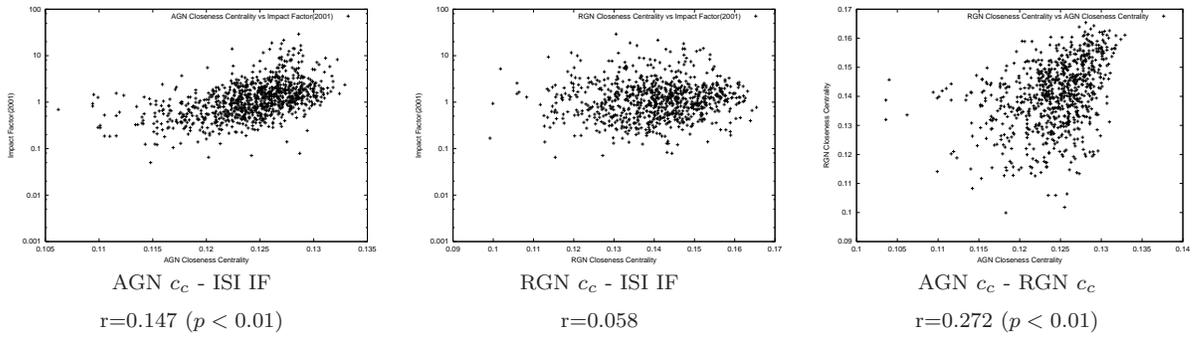

| AGN $c_c$ - ISI IF | RGN $c_c$ - ISI IF | AGN $c_c$ - RGN $c_c$ |
|---|---|---|
| r=0.147 ($p < 0.01$) | r=0.058 | r=0.272 ($p < 0.01$) |

Table 10
Comparison of closeness centrality for AGN and RGN, and the ISI IF these two subjects. Closeness and degree centrality values have non-significant correlations of respectively 0.058 and 0.030 to the ISI IF, indicating that although they seem to reflect a property of the LANL community of readers, they are not related to the ISI IF rankings for the same journals.

An overview of all centrality results is shown in Tables 9, 10, 11. In particular, we draw attention to the fact that the RGN and AGN centrality rankings do correlate. The AGN and RGN centrality rankings correlate at a 0.413 ($p < 0.01$) level, while the AGN and RGN closeness and betweenness centrality rankings correlate at lower, but statistically significant levels of 0.272 ($p < 0.01$) and 0.118 ($p < 0.01$). In other words, we find that none of the RGN centrality rankings correlate at significant levels with the ISI IF, but all are moderately related to the AGN centrality rankings.

| Rank | AGN Betweenness Centrality (AGN $c_b$) | RGN Betweenness Centrality (RGN $c_b$) | ISI IF |
|------|----------------------------------------|----------------------------------------|--------|
| 1 | P. IEEE | J. Chromatogr. A | Cell |
| 2 | Neural Networks | FEMS Microbiol Lett. | Curr. Opin. Cell Biol. |
| 3 | J. Theor. Biol. | J. Phys. D Appl. Phys. | Immunity |
| 4 | Physica D | Talanta | Trends Cell. Biol. |
| 5 | Chem. Geol. | J. Pharmaceut. Biomed. | Trends Neurosci. |
| 6 | Rep. Prog. Phys. | Photosynth. Res. | Trends Biochem. Sci. |
| 7 | Curr. Opin. Biotech. | Electron Lett. | Neuron |
| 8 | Curr. Opin. Chem. Biol. | Meas. Sci. Technol. | Surf. Sci. Rep. |
| 9 | Biosens. Bioelectron. | Chem. Phys. Lett. | Prog. Mater. Sci. |
| 10 | Geochim. Cosmochim. Ac. | Forensic Sci. Int. | Curr. Opinion Immunol. |

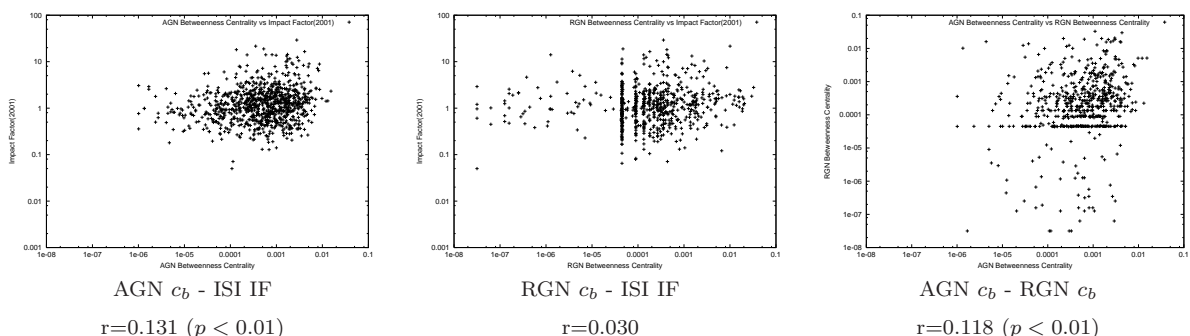

| AGN $c_b$ - ISI IF | RGN $c_b$ - ISI IF | AGN $c_b$ - RGN $c_b$ |
| r=0.131 ($p < 0.01$) | r=0.030 | r=0.118 ($p < 0.01$) |

Table 11
Comparison of betweenness centrality for AGN and RGN, and the ISI IF

## 7 Conclusion

Journal impact is an elusive concept which can be defined and determined in a number of different ways. We have provided a taxonomy of impact measures based on the distinction between author-based vs. reader-based data sets, and frequentist vs. structural metrics.

The ISI IF can be viewed as a common operationalization of the notion of journal impact, which we labeled $I_g$. The ISI IF's reliance on citation data, and its specific definition as a normalized citation frequency determined over the course of 2 years, implies a number of specific biases. Its use of citation frequencies clearly positions it in the domain of frequentist metrics which rely not so much on the structural position of an item in a network of relationships, but on the frequency with which the item has been preferred over others. In the case of the ISI IF, the community for which citation data has been collected is limited to scholarly authors and it therefore does not take into account readership or usage.

We have shown that valid networks of weighted journal relationships can be derived from reader downloads in a DL analyzed by applying the Retrieval Coherence Assumption. These networks, labeled Reader Generated Networks, are similar to citation graphs and other document networks such as the WWW, except that they have been derived from structural patterns of document downloads locally recorded in a DL's logs. The application of social network metrics to these reader-defined networks yields a measure of journal impact which is both structural and reader-defined. The resulting impact rankings reflect the local impact of journals, rather than the ISI IF which functions as a "global" measure of impact.

We have generated a similar network of weighted journal relationships from JCR citation data, labeled Author Generated Network, and applied the same structural metrics to this network. The rendered journal impact rankings were compared to those derived from the Reader Generated Network, and the ISI IF. Our results indicate the following: First, journal relationships in the RGN network seem to be valid and representative of the community whose downloads have shaped the network. Second, structural journal impact metrics derived from the RGN deviate strongly from the ISI IF. Third, the applied structural impact metrics correlate strongly with the ISI IF when calculated over the AGN, indicating they do validly operationalize journal impact, if we honor the assumption that the ISI IF does. Fourth, the AGN and RGN networks overlap to some degree, but exhibit striking differences.

This data suggests that the patterns by which a specific community of readers accesses documents induces a different, local, perspective of journal impact than the one provided by the common, global, operationalization of $I_g$: the ISI IF. From this, we may not conclude that the local metrics that were derived on the basis of download data gathered for a specific research community, the Los Alamos National Laboratory, are in fact an operationalization of $I_g$. The reason lies in the local nature of the RGN from which the metrics were derived, and the fact that we expect operationalizations of $I_g$ to have a global reach in order for them to be comparable and acceptable.

However, the widely-accepted ISI IF is computed on the basis of a global and representative sample of journals, across which citations patterns are counted. As was shown, other citation-related metrics, which effectively are other operationalizations of $I_g$, can be computed on the basis of the same RGN that results from this ISI-specific selection of journals. Similarly, one can easily imagine collecting usage information from a representative sample of research institutions worldwide, applying a procedure such as the proposed RCA to the resulting dataset, and in doing so obtaining an RGN with global properties.

This RGN need not be restricted to the selection of journals made by ISI, but it can include the wide variety of materials made available through DLs that are used during the research process. In fact, although our present analysis is focused on the creation of journal RGNs, the proposed methodology can be applied to other units of scholarly communication, such as articles, simulations, software and data sets (Sompel, Payette, Erickson, Lagoze, & Warner, 2004), given that download data is available. On the basis of the resulting RGNs, metrics can be computed that convey global, alternative operationalizations of $I_g$ for any particular units of scholarly communication.

This concept is attractive, especially since existing technologies such as the OAI-PMH (Lagoze, Sompel, Nelson, & Warner, 2002) can be used to harvest usage information from collaborating institutions. The OAI-PMH has already been applied to harvesting download logs in a restricted environment (Sompel, Young, & Hickey, 2003), and the log-gathering framework we envision could be based on the application of existing concepts to an open framework.

The framework we envision would enable the automatic creation of an RGN with global reach, for which various usage-based operationalizations of $I_g$ can be computed. It could be openly accessible to the scientific community, and it could yield new metrics for the evaluation of the performance of individuals, research groups, institutions and countries in the scholarly community. As such, it could balance the impact existing ISI-derived metrics have on these evaluation processes. As a result, it could eventually help to gnaw through the monopoly of scholarly publishers that is established by the necessity for researchers to publish in ISI-selected journals as a means to advance their careers. With the power-position of publishers diminished, a scholarly communication system can emerge in which information flows more openly.

**Acknowledgements**

This research has partly been funded by the Los Alamos National Laboratory Research Library as part of the Relationship Blackbox Modules project. We thank Somasekhar Vemulapalli (ODU) and Weining Xu (ODU) for their help in performing the data collection, normalization and analysis.

**Author Bios**

**Johan Bollen**: Johan Bollen is Assistant Professor at the Computer Science department of Old Dominion University, which he joined in 2002 after working for the Active Recommendations Project at the Los Alamos National Laboratory since 1999. He received his PhD in Experimental Psychology from the University of Brussels in 2001. His research focuses on the study of user behavior in digital information systems and its applications to scientometrics and recommender systems for digital libraries. His main interests are the application of social network theory to issues of DL impact ranking and the determination of research trends from large-scale usage patterns.

**Herbert Van de Sompel**: Herbert Van de Sompel graduated in Mathematics and Computer Science at Ghent University, and in 2000, obtained a Ph.D. there. For many years, he was Head of Library Automation at Ghent University. After having left Ghent in 2000, he has been Visiting Professor in Computer Science at Cornell University, and Director of e-Strategy and Programmes at the British Library. Currently, he is the team leader of the Digital Library Research and Prototyping Team at the Research Library of the Los Alamos National Laboratory. Herbert has played a major role in creating the Open Archives Protocol for Metadata Harvesting, the OpenURL Framework for Context-Sensitive Services, and the SFX linking server.

**Rick Luce**: Rick Luce is the Research Library Director at Los Alamos National Laboratory. He is an information technology pioneer internationally known for the cutting-edge digital library at Los Alamos. Rick was appointed Project Leader of the "Library Without Walls" digital library program in 1994 and he received a Los Alamos Distinguished Performance Award in 1996. The Library Without Walls was the first digital library program to deliver large-scale databases via the web (1994), interactive personal alerts (1995), and content linking (1996). Rick holds numerous digital library and electronic publishing positions, including Senior Advisor for Max Planck Society's Center for Information Management, the Executive Board of NISO, the UC Digital Media Innovations Program, and Course Director of the International School on the Digital Library and E-Publishing for Science and Technology in Geneva. He is a co-founder of the Open Archives Initiative and the Alliance for Innovation in Science and Technology Information consortium.

**Joan A. Smith**: Joan A. Smith is a doctoral student in computer science at Old Dominion University in Norfolk, Virginia. Ms. Smith began her doctoral studies at Old Dominion in 2002. She holds an M.A. degree in computer education from Hampton University in Hampton, Virginia, and a B.A. degree in natural science from the University of the State of New York in Albany. Her current research focuses on preservation of digital library resources for future access. She has been named the eighth recipient of the Zipf Fellowship in 2004.